\documentclass{article}
\usepackage{spconf,amsmath}
\usepackage[pdftex]{graphicx}
\usepackage{bm}
\usepackage{amsfonts}
\usepackage{cite}

\newcommand{\argmin}{\mathop{\rm arg\ min}\limits}

\title{Audio-Visual Speech Separation Using \\Cross-Modal Correspondence Loss}
  \name{Naoki Makishima, Mana Ihori, Akihiko Takashima, Tomohiro Tanaka, Shota Orihashi, Ryo Masumura}
	\address{NTT Media Intelligence Laboratories, NTT Corporation, Japan}
\begin{document}
\ninept
\maketitle
\begin{abstract}
We present an audio-visual speech separation learning method that considers the correspondence between the separated signals and the visual signals to reflect the speech characteristics during training.
Audio-visual speech separation is a technique to estimate the individual speech signals from a mixture using the visual signals of the speakers.
Conventional studies on audio-visual speech separation mainly train the separation model on the audio-only loss, which reflects the distance between the source signals and the separated signals.
However, conventional losses do not reflect the characteristics of the speech signals, including the speaker's characteristics and phonetic information, which leads to distortion or remaining noise.
To address this problem, we propose the cross-modal correspondence (CMC) loss, which is based on the cooccurrence of the speech signal and the visual signal.
Since the visual signal is not affected by background noise and contains speaker and phonetic information, using the CMC loss enables the audio-visual speech separation model to remove noise while preserving the speech characteristics.
Experimental results demonstrate that the proposed method learns the cooccurrence on the basis of CMC loss, which improves separation performance.
\end{abstract}
%
\section{Introduction}
\label{sec:intro}
Speech separation is a technique used to estimate the individual speech signals from an observed mixture of speech signals and is an important frontend step in wide range of tasks such as automatic speech recognition~\cite{YIsik2016_ASRSEusingDC,KKinoshita2020_ASRSE} and speaker diarization~\cite{QWang2018_speakerDiarizationLSTM,NKanda2019_ASRSD}.
In single-channel speech separation, fully neural network based techniques such as deep clustering~\cite{JHershey2016_deepclustering,YIsik2016_ASRSEusingDC} and permutation invariant training (PIT)~\cite{DYu2017_PIT,MKolbak2017_uPIT} have shown promising performance.

In most of the audio-only studies, the individual speech signals are extracted from a mixture of signals without using any information about the speaker.
This leads to the inherent ambiguity in speaker labeling of the separated signals, which is referred to as the permutation problem.
Although many studies have addressed this problem~\cite{JHershey2016_deepclustering,DYu2017_PIT}, they still cannot assign the correct speaker labels to the separated signals with similar vocal characteristics~\cite{RLiu2019_AVDeepclustering}.
On the other hand, recent studies have examined audio-visual speech separation~\cite{AEphrat2018_LookingtoListen,AGabbay2018_SeeingthroughNoise,RLiu2019_AVDeepclustering,CLi2020_DeepAVSSwithAttention} and enhancement~\cite{AGabbay2018_VSE,TAfouras2018_conversation,TOchiai2019_speakerbeam,JHou2018_AVSEDCNN,WWang2020_RobustSE} that use the visual information as auxiliary input to the audio-visual speech separation model.
The most common visual information used in these studies is facial movements of the speaker.
The advantage of using facial movements is that the correspondence between facial movements and speech signals can be explicitly used.
This correspondence should be beneficial for speech separation because it has been shown that viewing a speaker's face enhances a human listener's understanding of the speaker's speech~\cite{EGolumbic2013_cocktail}.
Moreover, using visual signals enables the audio-visual separation model to extract the corresponding speech of the speaker in the video.
Therefore, in audio-visual speech separation, the permutation problem is automatically solved.
In practice, such visual signals are  available in applications such as video conferencing and human-machine interaction,
and several studies have shown that audio-visual speech separation outperforms audio-only speech separation~\cite{AEphrat2018_LookingtoListen,AGabbay2018_SeeingthroughNoise,RLiu2019_AVDeepclustering,CLi2020_DeepAVSSwithAttention}.

Many types of architectures and strategies for audio-visual speech separation have been proposed.
For example, in~\cite{AGabbay2018_SeeingthroughNoise}, speech spectrograms converted from silent video by  \textit{vid2speech}~\cite{AEphrat2017_vid2speech} are used as prior knowledge for creating time-frequency (T-F) masks, and in~\cite{RLiu2019_AVDeepclustering}, audio and visual features are separately fed into the networks, which are then concatenated to perform T-F clustering.
These studies successfully combine visual signals and  mixture audio.
However, a limitation of these studies is that they train the audio-visual speech separation model on an audio-only loss function such as the mean squared error (MSE) loss and L1 loss of the separated signals and the source signals, which does not guarantee the speaker's characteristics or the appropriate utterance in the separated signals.
Thus, the separated signals can contain distortion or interference as reported in~\cite{PLoizou2011_msedistortion,SFu2018_pesqstoiloss}.
We have addressed this problem, and our key idea is to explicitly consider the relationship between the separated signals and the corresponding visual signals.
Specifically, we assume two things.
First, the speech characteristics such as the speaker's characteristics and phonetic information in the completely separated signal correspond to those in the target speaker's visual signal.
Second, this correspondence does not exist when the separated signal contains noise or the speaker differs from the one in the visual signal.

In this paper, we present a new audio-visual speech separation learning method based on the above assumptions that considers the correspondence between the visual signals and the separated signals.
This is enabled by the introduction of the new cross-modal correspondence (CMC) loss into the audio-visual speech separation model.
The CMC loss is calculated using the cosine similarity between the visual features and the separated signal's features, where the similarity between the visual signal and the corresponding separated signal is maximized while the similarity between the visual signal and a different speaker's separated signal is minimized.
Since the visual signal is not affected by background noise and contains speaker and phonetic information, the loss should be beneficial in removing noise while preserving the characteristics of the target speaker's speech.

Although many studies have examined the correspondence between audio signals and visual signals~\cite{YAytar2016_Soundnet,Dharwath2016_UnsupervisedLearningofSpokenLanguagewithVideo,RArandjelovic2017_LookListenLearn,HZhao2018_SoundofPixels,AOwens2018_AVsceneAnalysis}, they mainly focus on the correspondence between video and the object's sound such as musical instruments and tools.
To the best of our knowledge, ours is the first study that trains an audio-visual speech separation model to consider the characteristics of the separated speech signals and the visual signals of the speakers.
Experimental results using the Lip Reading Sentences 3 dataset (LRS3-TED)~\cite{TAfouras2018_LRS3} demonstrate that the proposed method using CMC loss captures the correspondence between the visual signals and the separated signals, which improves separation performance.

\section{Audio-Only Speech Separation}
\label{sec:related}
\subsection{Basic Framework}
Let $N$ be the number of speakers.
The short time Fourier transforms (STFTs) of the source, observed, and estimated signals are defined as $\bm S_n \in \mathbb{C}^{I\times J}$, $\bm X \in \mathbb{C}^{I\times J}$, and $\bm Y_n \in \mathbb{C}^{I\times J}$, respectively, where $n = 1,\ldots, N$ denotes the index of the speaker, $I$ denotes the number of frequency bins, and $J$ denotes the number of time frames.
The goal of speech separation is to recover $\bm S_n$ from mixture $\bm X$.

In fully neural network based separation, a common strategy is to use the network to estimate the T-F masks~\cite{DWang2018_DNNSSoverview}.
We denote $\mathcal F(\cdot)$ as the neural network based function used to estimate the T-F masks.
The set of estimated signals $\bm Y =\{\bm Y_1,\ldots,\bm Y_N\}$ is obtained as
\begin{align}
  \bm M &= \mathcal F(|\bm X|; \theta),\label{eq:audio-onlyDNN}\\
  \bm Y &= g(\bm M) \odot \bm X \label{eq:audio-onlymask},
\end{align}
where $|\cdot|$ denotes element-wise absolute operation, $\theta$ is the set of network parameters, $g$ is the function to constrain $\bm M$ between $0$ and $1$, and $\odot$ is the element-wise product.

\subsection{Permutation Invariant Training}
In audio-only speech separation, $\bm Y_n$ is estimated from mixture $\bm X$ by using \eqref{eq:audio-onlyDNN} and \eqref{eq:audio-onlymask} without any speaker information.
Moreover, in the speaker-independent speech separation task, $\mathcal F$ is trained to extract an arbitrary speaker's speech from the mixture.
Thus, the correspondence between $\bm Y_{n'}$ and $\bm S_n$ w.r.t. $n'$ and $n$ is unknown, which is a problem that must be solved in order to achieve appropriate training.
This problem is referred to as the permutation problem.
In utterance-level PIT (uPIT)~\cite{MKolbak2017_uPIT}, the following loss function is used to address the permutation problem:
\begin{align}
  L_{\mathrm{uPIT}} = \frac{1}{IJN}\sum_n\||\bm Y_n| - |\bm S_{\phi^*(n)}|\odot\cos(\psi_X-\psi_{\phi^*(n)})\|_F^2,\label{eq:uPIT}
\end{align}
where $\|\cdot\|_F$ denotes the Frobenius norm of the matrix, $\psi_X$ and $\psi_{\phi^*(n)}$ denote the phases of mixture $\bm X$ and source $\bm S_{\phi^*(n)}$, respectively, $\phi^*(n)$ is the permutation that minimizes the utterance-level separation error and is defined as
\begin{align}
  \phi^* = \argmin_{\phi\in\mathcal{P}}\sum_n\||\bm Y_n| - |\bm S_{\phi(n)}|\odot\cos(\psi_X-\psi_{\phi(n)})\|_F^2,
\end{align}
and $\mathcal{P}$ is the set of all $N!$ permutations.
The idea of uPIT is to calculate the MSE loss for all possible $N!$ assignments and set the permutation in accordance with the lowest MSE loss.

\section{Proposed Audio-Visual Speech Separation~Method}
\label{sec:proposed}
\subsection{Strategy}
\begin{figure}
  \begin{center}
  \includegraphics[width=1.0\columnwidth]{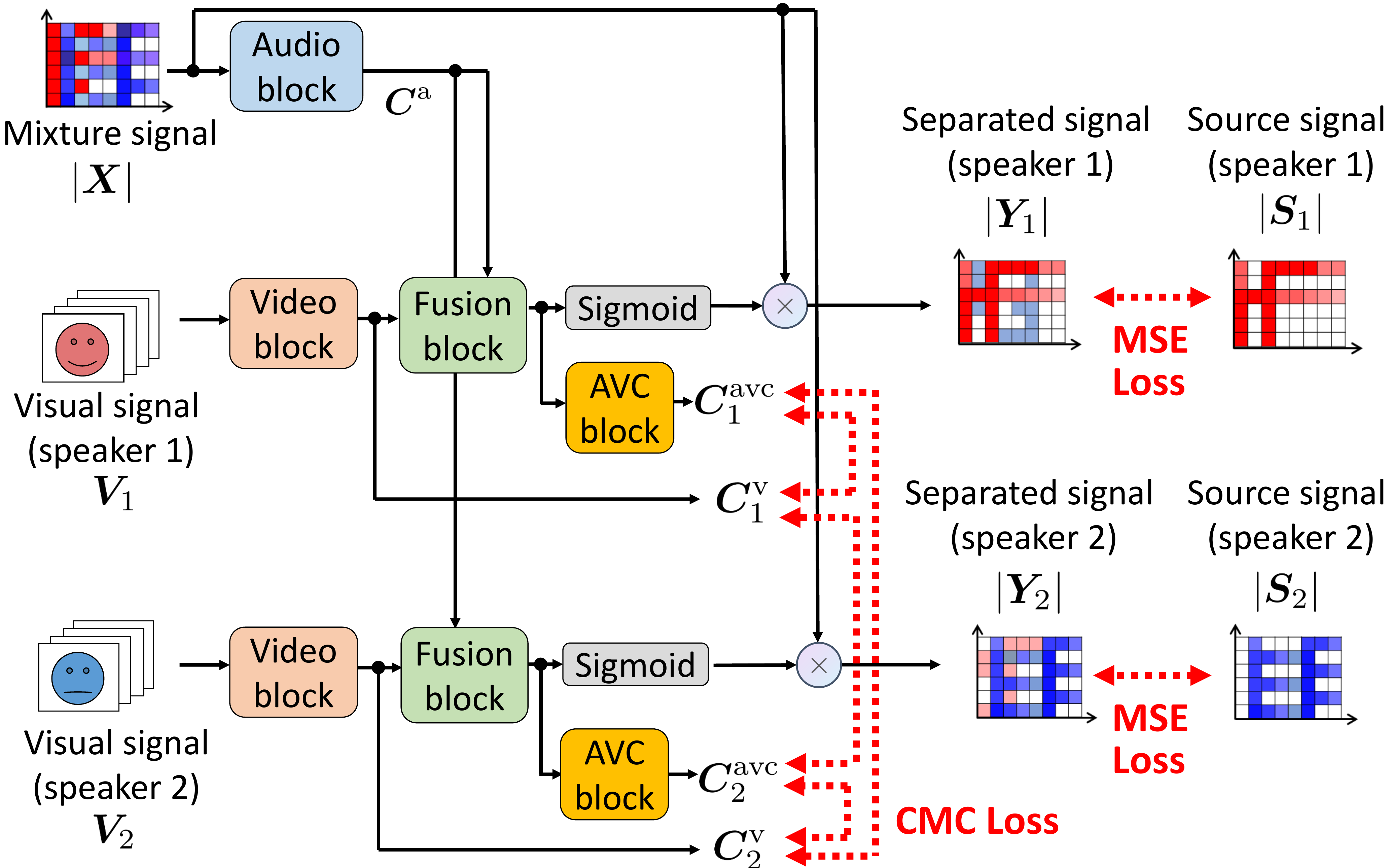}
  \vspace{-20pt}
  \end{center}
  \caption{Overview of proposed method with $N=2$. Blocks with same label share network parameters.}
  \vspace{-10pt}
  \label{fig:overview}
\end{figure}
In most audio-visual speech separation studies, the separation model is trained to minimize the audio-only loss function such as the MSE loss and L1 loss of the separated signal and the source signal.
However, as described in section~\ref{sec:intro}, a separation model trained to minimize these losses does not consider the characteristics of the speech signal, which leads to distortion or remaining noise in the separated signals.
Our proposed method addresses this problem by using CMC loss, which considers the cosine similarity between the feature vector of the separated signal and the visual signal.

As shown in Fig. \ref{fig:overview}, the proposed method estimates $\bm Y_n$ and calculates the CMC loss using four types of blocks: an audio block, and, for each speaker, a video block, an audio-visual fusion block, and an audio-visual correspondence (AVC) block.
All the blocks share the same network parameters across speakers, which enables the same framework to handle arbitrary $N$ speech separation and reduces the number of model parameters.
The audio and video blocks learn the representations of the mixture and visual signals, respectively.
Then, each audio-visual fusion block separates the mixture representation using the visual representation received from the corresponding video block.
Next, each AVC block extracts the feature of the separated signal received from the corresponding fusion block.
Note that since the AVC block is not required to output $\bm Y_n$ as illustrated in Fig.~\ref{fig:overview}, the AVC block is removed from the proposed method after training.
Thus, the computational cost of the propose method is the same as that of the conventional audio-visual speech separation methods during inference.
The modeling of each block type is described in section~\ref{sec:modeling}.

We denote the output of the audio, video, and AVC blocks as
$\bm C^{\mathrm a}\in\mathbb{R}^{D\times J}$, $\bm C_n^{\mathrm v}\in\mathbb{R}^{D\times J}$, and $\bm C_n^{\mathrm {avc}}\in\mathbb{R}^{D\times J}$, respectively, where $D$ denotes the dimension of the feature vector.
We call the pair of $\bm C_{n'}^{\mathrm v}$ and $\bm C_n^{\mathrm {avc}}$ a negative pair when $n' \not = n$ and call it a positive pair when $n' = n$.
Since $\bm C_n^{\mathrm v}$ is used to extract the $n$th speaker's speech signal from $\bm C^{\mathrm {a}}$ in an audio-visual fusion block, we assume that $\bm C_n^{\mathrm v}$ contains the target speech information such as the speaker's characteristics and phonetic information.
If noise or distortion remains in the output of an audio-visual fusion block, there are mainly two problems.
\begin{itemize}
  \item The block fails to extract the corresponding audio feature of $\bm C_n^{\mathrm v}$ from $\bm C^{\mathrm a}$.
  \item The visual block output $\bm C_n^{\mathrm v}$ is not learned as an appropriate feature for extracting the $n$th speaker.
\end{itemize}
Using the proposed CMC loss overcomes both problems.
CMC loss is composed of two terms; one to maximize the cosine similarity of the positive pairs and one to minimize the cosine similarity of the negative pairs, which is formulated in section~\ref{sec:cmc}.
Maximizing the cosine similarity of the positive pairs results in the audio-visual fusion blocks being trained to separate signals containing the feature included in $\bm C_n^{\mathrm v}$, which helps remove noise and distortion in the separated signals.
Minimizing the cosine similarity results in $\bm C_{n'}^{\mathrm v}$ and $\bm C_n^{\mathrm {avc}}  (n' \not= n)$ being trained to be orthogonal.
Since $\bm C_{n}^{\mathrm v}$ and $\bm C_n^{\mathrm {avc}}$ are trained to maximize the cosine similarity, $\bm C_{n'}^{\mathrm v}$ and $\bm C_{n}^{\mathrm v}$ are also trained to be orthogonal.
This should help in extracting the target speaker's speech in the audio-visual fusion block.
The implementation of each block type is described in section \ref{sec:implementation}.

\subsection{Modeling\label{sec:modeling}}
In this section, we formulate the audio-visual speech separation model used to estimate $\bm Y_n$ and calculate the CMC loss.
Let $\bm V_n\in\mathbb{R}^{(W\times H\times C)\times F}$ be the visual signal, where $W$ and $H$ denote the width and height of the cropped image of the speaker's face, respectively, $C$ denotes the number of channels, and $F$ denotes the number of frames in a video.
Audio mixture $\bm X$ and visual signal $\bm V_n$ are separately fed into networks to obtain mixture representation $\bm C^{\mathrm{a}}$ and visual representation $\bm C^{\mathrm{v}}_n$:
\begin{align}
  \bm C^{\mathrm a} &= \mathrm{AudioBlock}(|\bm X|;\theta^{\mathrm a}),\\
  \bm C^{\mathrm v}_n &= \mathrm{VideoBlock}(\bm V_n;\theta^{\mathrm v}),
\end{align}
where $\mathrm{AudioBlock}(\cdot)$ denotes the audio block, $\theta^{\mathrm{a}}$ denotes the parameter of the audio block, $\mathrm{VideoBlock}(\cdot)$ denotes the video block, and $\theta^{\mathrm{v}}$ denotes the parameter of the video block.
Given $\bm C^{\mathrm{a}}$ and $\bm C^{\mathrm{v}}_n$, an audio-visual fusion block outputs the separated signal as
\begin{align}
  \bm M_n &= \mathrm{FusionBlock}(\bm C^{\mathrm{a}}, \bm C^{\mathrm{v}}_n; \theta^{\mathrm{f}}), \label{eq:mask}\\
  \bm Y_n &= g(\bm M_n) \odot \bm X,
\end{align}
where $\mathrm{FusionBlock}(\cdot)$ denotes the audio-visual fusion block and $\theta^{\mathrm{f}}$ denotes its parameter.
As a parallel path for $\bm M_n$, the AVC block outputs the feature of the separated signal as
\begin{align}
  \bm C_n^{\mathrm {avc}} = \mathrm{AVCBlock}(\bm M_n; \theta^{\mathrm{avc}}), \label{eq:sepfeature}
\end{align}
where $\mathrm{AVCBlock}(\cdot)$ denotes the AVC block and $\theta^{\mathrm{avc}}$ denotes its parameter.

\subsection{Training\label{sec:cmc}}
In this section, we formulate the proposed CMC loss.
We define the CMC loss as
\begin{align}
  L_{\mathrm {CMC}} &= \sum_n\sum_j\biggr[\sum_{n' \not = n}\left|d(\bm c_{nj}^{\mathrm v}, \bm c_{n'j}^{\mathrm {avc}})\right| - d(\bm c_{nj}^{\mathrm v}, \bm c_{nj}^{\mathrm {avc}})\biggl],\label{eq:avc}\\
  d(\bm a, \bm b) &= \frac{
      \bm a^{\mathrm T} \bm b
      }{
    \|\bm a \|\|\bm b \|
    },
\end{align}
where $\bm c_{nj}^{\mathrm v}$ and $\bm c_{nj}^{\mathrm {avc}}$ denote the $j$th column vector of $\bm C_{n}^{\mathrm v}$ and $\bm C_{n}^{\mathrm {avc}}$, respectively, $^{\mathrm T}$ denotes the transpose, and $\|\cdot\|$ denotes the L2 norm of the vector.
The first term in \eqref{eq:avc} minimizes the cosine similarity of the negative pairs, and the second term maximizes the cosine similarity of the positive pairs.
Compared with the speech enhancement technique using audio speaker clues~\cite{Mdelcroix2018_speakerbeam}, where  pre-recorded speaker's speech is summarized over the time frame as a time-invariant characteristic of the speaker, we use the frame-wise characteristics of the separated signals and the visual signals in order to use the phonetic information in the visual signals.

The CMC loss combined with the conventional MSE loss constitutes the loss function of the proposed method:
\begin{align}
  L_{\mathrm{proposed}} &= L_{\mathrm{MSE}} + \lambda L_{\mathrm {CMC}}, \label{eq:prop}\\
  L_{\mathrm{MSE}} &= \frac{1}{IJN}\sum_n\||\bm Y_n| - |\bm S_n|\|_F^2,\label{eq:mse}
\end{align}
where $\lambda$ is the loss weight.
Note that \eqref{eq:prop} is the conventional MSE loss when $\lambda = 0$, which is the baseline used to evaluate the proposed method~\cite{AEphrat2018_LookingtoListen,AGabbay2018_VSE}.

\subsection{Implementation\label{sec:implementation}}
In this section, we describe each type of block in detail.
The input to $\mathrm{AudioBlock}(\cdot)$ is the magnitude spectrogram of the speech mixture.
The audio block consists of one fully connected layer and three bidirectional long short-term memory (BLSTM) layers, which capture the long-term dependencies of the speech signal.
The forward and backward outputs of BLSTM are concatenated, and each BLSTM output is followed by tanh activation and dropout of 0.5.

The network $\mathrm{VideoBlock}(\cdot)$ takes as input a cropped image of the speaker's face, which is converted to grayscale.
The combination of a 3D convolution layer and 18-layer ResNet~\cite{KHe2016_ResNet} with a squeeze-and-excitation (SE)~\cite{JHu2020_SENet} module is used as the video block, similar to that used in a previous study~\cite{TAfouras2018_conversation}.
For 3D convolution, we use a $5\times 7\times 7$ kernel with $(2, 3, 3)$ padding, and $(1, 1, 1)$ stride to obtain output the same size as the input.
The architecture of the ResNet is the same as that described in~\cite{KHe2016_ResNet} except that all the residual connections are replaced by the SE-module~\cite{JHu2020_SENet}.
We set the reduction ratio of the SE-module to 16.
If a shortcut connection causes a mismatch in the dimensions, zero-padding is used to make them match.
The output of the ResNet is passed to a 2D global average pooling layer and a BLSTM layer.
Since the number of audio frames $J$ does not match that of video frames $F$ due to a gap in the sampling frequency, the corresponding frames must be aligned.
In this implementation, the output of a video block is repeated for several audio frames to make them match.

The network $\mathrm{FusionBlock}(\cdot)$ consists of a $1\times 1$ convolution layer with one output channel, one BLSTM layer, and one linear layer.
In the fusion block, $\bm C^{\mathrm{a}}$ and $\bm C^{\mathrm{v}}_n$ are concatenated along the channel dimension, which is then fed into the convolution layer.
The sigmoid function is adopted as $g(\cdot)$.
The network $\mathrm{AVCBlock}(\cdot)$ consists of one linear layer followed by tanh activation.

\section{Evaluation}
\label{sec:experiments}
\subsection{Datasets}
We evaluated the proposed method by conducting a speech separation task.
The model was trained on the LRS3-TED dataset~\cite{TAfouras2018_LRS3}, which consists of thousands of spoken sentences from TED and TEDx videos.
We prepared two training datasets: a 40-hour mixture dataset and an 80-hour mixture dataset containing sentences from 850 speakers.
The validation dataset was a 2-hour mixture dataset containing sentences from 500 speakers, and the test dataset was a 5-hour mixture dataset containing sentences from 412 speakers.
The speakers of the training, development, and test dataset were randomly chosen from \textit{pretrain}, \textit{trainval}, and \textit{test} sets in the LRS3-TED dataset, respectively.

We created two-speaker mixtures by mixing the utterances of different speakers at a signal-to-noise ratio between 0 and 5 dB.
We downsampled the sampling frequency of the audio signal to 8 kHz and halved the frame rate of the video to reduce the computational cost.

\subsection{Settings}
We compared the performance of the proposed method with that of uPIT~\cite{MKolbak2017_uPIT} (an audio-only method) and that of a conventional audio-visual method that uses the MSE loss as a loss function (AV baseline)~\cite{AEphrat2018_LookingtoListen}.
The uPIT architecture and the training details were the same as the ones described in~\cite{MKolbak2017_uPIT}.
For the AV baseline, we created the same architecture as that of the proposed method.
The AV baseline was trained without the CMC loss ($\lambda = 0$ in \eqref{eq:prop}).
An appropriate value for $\lambda$ in \eqref{eq:prop} was determined experimentally, as described in section~\ref{sec:results}.
All the blocks in the AV baseline and the proposed method were trained simultaneously from scratch.

An STFT was performed using a 400-ms-long Hamming window with a 200-ms-long shift to convert the audio signals into T-F domain signals.
The audio-visual speech separation model was optimized using the Adam algorithm~\cite{DKingma2015_Adam} with a minibatch size of 8.
We set the initial learning rate of the algorithm to $2\times 10^{-5}$.
The training steps were stopped if the loss on the validation set does not decrease for 20 epochs in succession.
We used the signal-to-distortion ratio (SDR)~\cite{EVincent2006_SDR}, perceptual evaluation of speech quality (PESQ)~\cite{ARix2001_PESQ}, and short-time objective intelligibility (STOI)~\cite{CTaal2011_STOI} as metrics to evaluate total separation performance.

\subsection{Correspondence Between Separated Signals and Visual Signals\label{sec:results2}}
\begin{figure}
  \begin{center}
  \includegraphics[width=1.0\columnwidth]{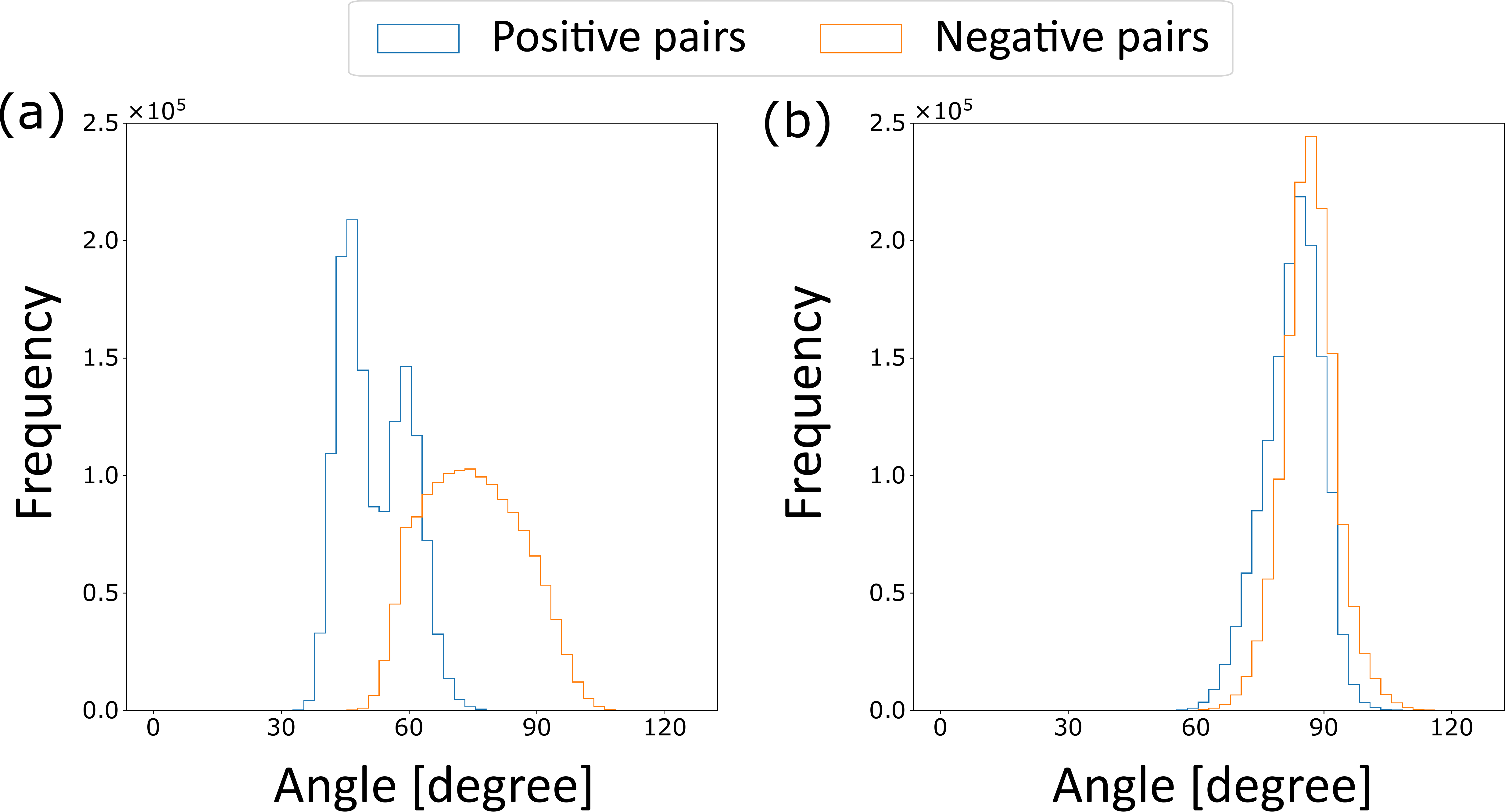}
  \vspace{-20pt}
  \end{center}
  \caption{Histogram of angles formed by $\bm c_{nj}^{\mathrm v}$ and $\bm c_{nj}^{\mathrm {avc}}$.
  Blue line represents histogram of positive pairs and orange line represents that of negative pairs for (a) proposed method and (b) AV baseline.}
  \vspace{-10pt}
  \label{fig:histogram}
\end{figure}

To examine whether the audio-visual speech separation model learns the audio-visual correspondence by using the CMC loss, we compared the histogram of the angles formed by $\bm c_{nj}^{\mathrm v}$ and $\bm c_{n'j}^{\mathrm {avc}}$ using the test dataset.
Since the AV baseline does not have the learned AVC block, we used the AVC block of the proposed method to obtain $\bm c_{nj}^{\mathrm {avc}}$ for the AV baseline.
Fig.~\ref{fig:histogram} (a) shows the histogram after training for the proposed method ($\lambda=1$)  and Fig.~\ref{fig:histogram} (b) shows the histogram after training for the AV baseline, where the blue line represents the positive pairs and the orange line represents the negative pairs.
From Fig.~\ref{fig:histogram} (a), we can see that the positive pairs and the negative pairs formed different distributions in the proposed method, which means that the proposed method can distinguish the different speaker's characteristics in the separated signals and the visual signals.
On the other hand, with the AV baseline method, the positive pairs and the negative pairs formed almost the same histogram, as illustrated in Fig.~\ref{fig:histogram} (b).
This demonstrates that the proposed method using CMC loss captures the frame-wise correspondence between the separated signals and the visual signals, which cannot be captured by the AV baseline using the MSE loss.

\subsection{Separation Performance\label{sec:results}}
Table \ref{table:res_40h} shows the experimental results of unprocessed mixture, uPIT, the AV baseline,
and the proposed method for the 40-hour mixture training dataset.
We set $\lambda$ to $(0.1, 1, 10, 100)$ and examined the separation performance of the proposed method.
The results show that the proposed method outperformed the AV baseline for all the metrics.
Specifically, the proposed method with $\lambda=1$ achieved the best separation performance.
\begin{table}[htb]
  \caption{Evaluation results for 40-hour mixture training dataset.}
  \begin{center}
  \begin{tabular}{l|ccc} \hline
    \textbf{Method} & \textbf{SDR} & \textbf{PESQ} & \textbf{STOI} \\ \hline
    Mixture & 0.49 & 1.60 & 0.683 \\
    uPIT & 6.67 & 1.94 & 0.777 \\
    AV baseline & 7.29 & 2.14 &  0.817\\
    Proposed method $(\lambda=0.1)$& 7.75 & 2.18 & 0.825\\
    Proposed method $(\lambda=1)$& \textbf{7.77} & \textbf{2.19} & \textbf{0.827}\\
    Proposed method $(\lambda=10)$& 7.59 & 2.14 & 0.819\\
    Proposed method $(\lambda=100)$& 7.48 & 2.13 & 0.817\\ \hline
  \end{tabular}
  \label{table:res_40h}
  \end{center}
  \vspace{-20pt}
\end{table}

Table \ref{table:res_80h} shows the experimental results for the 80-hour mixture training dataset.
Since the test data were the same as those used to obtain Table \ref{table:res_40h}, the results for the mixture signal are omitted from Table \ref{table:res_80h}.
The results show that the proposed method outperformed the AV baseline even when the amount of data was increased.
Specifically, the proposed method achieved better PESQ and STOI score than the AV baseline.
This indicates that the proposed method using CMC loss preserves the speech characteristics and removes distortion that affects  intelligibility and quality of the speech signals by considering the correspondence between the visual signals and the separated signals.
\begin{table}[htb]
  \begin{center}
  \caption{Evaluation results for 80-hour mixture training dataset.}
  \begin{tabular}{l|ccc} \hline
    \textbf{Method} & \textbf{SDR} & \textbf{PESQ} & \textbf{STOI} \\ \hline
    uPIT & 7.12 & 1.97 & 0.789 \\
    AV baseline & 8.46 & 2.27 & 0.843 \\
    Proposed method $(\lambda=1)$ & \textbf{8.85} & \textbf{2.39} & \textbf{0.854} \\ \hline
  \end{tabular}
  \label{table:res_80h}
  \end{center}
  \vspace{-20pt}
\end{table}

\section{Conclusions}
\label{sec:conclusion}
Our proposed audio-visual speech separation method considers the correspondence between the separated signals and the visual signals to reflect the speech characteristics during training.
It uses the new CMC loss to remove distortion and remaining noise.
The CMC loss is calculated on the basis of the cosine similarity between the features of the visual signals and the separated signals.
The similarity of positive pairs is maximized while that of negative pairs is minimized.
Since the visual signal is not affected by background noise and contains speaker and phonetic information, the CMC loss enables the model to remove noise while preserving the characteristics of the speech signal.
Experiments using the LRS3-TED dataset demonstrated that introducing the CMC loss enables the audio-visual speech separation model to learn the correspondence between the visual signals and the separated signals, which improves separation performance.

\bibliographystyle{IEEEbib}

\end{document}